\documentclass[conference]{IEEEtran}

\makeatletter

\usepackage{blindtext}
\usepackage{eso-pic}
\IEEEoverridecommandlockouts

\usepackage{cite}
\usepackage{amsmath,amssymb,amsfonts}
\usepackage{algorithmic}
\usepackage{graphicx}
\usepackage{textcomp}
\usepackage{xcolor}
\def\BibTeX{{\rm B\kern-.05em{\sc i\kern-.025em b}\kern-.08em
    T\kern-.1667em\lower.7ex\hbox{E}\kern-.125emX}}

\usepackage{eso-pic}

\begin{document}
\title{\vspace*{1cm} Quantixar: High-Performance Vector Data Management System\\
}

\author{\IEEEauthorblockN{Gulshan Yadav}
\IEEEauthorblockA{\textit{Department of Information Technology} \\
\textit{A. P. Shah Institute of Technology}\\
Thane, Maharashtra \\
gulshanyadav@apsit.edu.in}
\and
\IEEEauthorblockN{RahulKumar Yadav}
\IEEEauthorblockA{\textit{Department of Information Technology} \\
\textit{A. P. Shah Institute of Technology}\\
Thane, Maharashtra \\
20104093@apsit.edu.in}
\and
\IEEEauthorblockN{Mansi Viramgama}
\IEEEauthorblockA{\textit{Department of Information Technology} \\
\textit{A. P. Shah Institute of Technology}\\
Thane, Maharashtra \\
20104115@apsit.edu.in}
\and
\IEEEauthorblockN{Mayank Viramgama}
\IEEEauthorblockA{\textit{Department of Information Technology} \\
\textit{A. P. Shah Institute of Technology}\\
Thane, Maharashtra \\
20104119@apsit.edu.in}
\and
\IEEEauthorblockN{Apeksha Mohite}
\IEEEauthorblockA{\textit{Department of Information Technology} \\
\textit{A. P. Shah Institute of Technology}\\
Thane, Maharashtra \\
atmohite@apsit.edu.in}
}

\maketitle
\begin{abstract}
Traditional database management systems need help efficiently represent and querying the complex, high-dimensional data prevalent in modern applications. Vector databases offer a solution by storing data as numerical vectors within a multi-dimensional space. This enables similarity-based search and analysis, such as image retrieval, recommendation engine generation, and natural language processing. This paper introduces Quantixar, a vector database project designed for efficiency in high-dimensional settings. Quantixar tackles the challenge of managing high-dimensional data by strategically combining advanced indexing and quantization techniques. It employs HNSW indexing for accelerated ANN search. Additionally, Quantixar incorporates binary and product quantization to compress high-dimensional vectors, reducing storage requirements and computational costs during search. The paper delves into Quantixar’s architecture, specific implementation, and experimental methodology.
\end{abstract}



\section{Introduction}
The explosion of data produced by modern applications, from image repositories to text corpora, presents challenges for traditional database management systems. These systems, designed primarily for structured data, need help with the complexity and high dimensionality inherent in representations commonly used in machine learning and artificial intelligence tasks. Vector databases provide a compelling solution. Vector databases unlock similarity-based search by representing data points as numerical vectors within a multi-dimensional space. This capability revolutionizes tasks such as image retrieval, recommendation engine generation, and natural language processing.
This paper introduces Quantixar, a vector database project purpose-built for efficiency and scalability. Quantixar tackles the critical challenge of managing high-dimensional data, focusing on advanced indexing and quantization techniques. At its core lies the utilization of HNSW (Hierarchical Navigable Small World) indexing, a robust method for approximate nearest neighbor (ANN) search, crucial for accelerating similarity queries \cite{MalkovYashunin2016ANNHierarchicalNavigableGraphs, wang2021comprehensive}. To further optimize performance, Quantixar incorporates binary and product quantization. These techniques compress high-dimensional vectors, reducing storage requirements and computational costs during search processes. The optimizations made by Quantixar are optional and can be configured by a user based on his requirements. 
The effectiveness of vector databases hinges on their ability to handle the ``curse of dimensionality." As the number of dimensions in a dataset increases, the efficiency of similarity searches degrades\cite{zhong2015efficient, jin2024curator}. Quantixar's design addresses this challenge by employing cosine similarity for distance calculations as a default method. This technique has proven to be more resilient to the curse of dimensionality, where the effectiveness of Euclidean distance degrades as the number of dimensions increases \cite{altman2018curse}.
The remainder of this paper delves deeper into Quantixar's architecture, its specific implementation of indexing and quantization methods, and the experimental methodology used with the ANN Benchmark. We present a thorough analysis of the results, highlighting the strengths and potential areas for future development in the Quantixar system.

\section{Background}

\subsection{Curse of Dimensionality}

In vector databases and high-dimensional data analysis, the ``curse of dimensionality" represents a collection of phenomena that degrade the performance and meaningfulness of traditional distance-based similarity searches as the number of dimensions increases\cite{altman2018curse}. Understanding this curse is crucial, as it highlights the core challenge systems like Quantixar are designed to address.

\subsubsection{Theoretical Underpinning}
Consider a scenario where data points are randomly distributed within a unit hypercube (a cube where each side has a length of 1). As the dimensionality ``d" of the space increases, the volume of this hypercube grows exponentially. Imagine a smaller hypercube inscribed within the larger one, with its sides being some fraction 'r' of the outer hypercube's sides. Surprisingly, as dimensionality increases, even a substantial fraction like r=0.8 will cause the inner hypercube's volume to shrink towards zero relative to that of the outer cube.

The consequence is that in high-dimensional spaces, the vast majority of the volume is concentrated within a thin 'shell' near the boundaries of the space. This leads to a counterintuitive result:  the distances between randomly distributed points tend to become very similar. Mathematically, the variance of these distances decreases, making it increasingly difficult to discriminate between what we would intuitively consider ``near" and ``far" neighbors.

\subsubsection{Implications for Search Efficiency}
Traditional distance metrics such as Euclidean distance, which many algorithms like KNN (k-nearest neighbors) rely on, lose their effectiveness in high dimensions\cite{kouiroukidis2011effects}. Imagine trying to find the nearest neighbors to a query point. As dimensionality increases, the search process needs to cover an exponentially growing volume of space while the distances between points become less informative. This combination leads to several challenges:
\begin{itemize}
    \item 
    \textbf{Computational Cost:} Nearest neighbor search becomes computationally demanding, potentially rendering real-time similarity-based tasks infeasible.
    \item
    \textbf{Data Sparsity:} The volume of space proliferates that even large datasets become sparse, hindering the discovery of meaningful relationships.
\end{itemize}

\subsection{Approaches to Address High-Dimensionality Challenges}
\subsubsection{Hierarchical Navigable Small World (HNSW)}
HNSW is a graph-based indexing technique renowned for its efficiency in approximate nearest neighbor (ANN) search with high dimensional spaces. It excels in scenarios where exact nearest neighbor results can be sacrificed for significantly improved query speeds\cite{wang2021comprehensive}. The core principles behind HNSW are:
\begin{itemize}
    \item \textbf{Graph Construction:} HNSW builds a multi-layered graph. Links between graph nodes are established based on proximity and maintaining a "small-world" property, with long-range connections between nodes that might be distant but share similar neighborhoods. It avoids the need to exhaustively explore every local neighborhood, which would be time-consuming in high dimensions\cite{wang2021comprehensive}.
    \item \textbf{Hierarchical Structure:} Lower graph layers contain increasingly refined subsets of the dataset. This hierarchy allows search algorithms to narrow down relevant portions of the data space quickly\cite{wang2021comprehensive}.
    \item \textbf{Greedy Traversal:} During a query, HNSW employs a greedy search algorithm that traverses the graph. Starting at the top, it determines which general region of the space is most likely to contain the query's nearest neighbors. With each layer down, it zeroes in on an increasingly specific subset of data points. The hierarchy lets HNSW rapidly discard vast, irrelevant portions of the dataset, improving efficiency\cite{wang2021comprehensive}.
\end{itemize}

\subsubsection{Quantization: Compressing Vectors for Efficiency}
Quantization methods strategically reduce the precision of high-dimensional vectors, leading to significant gains in storage efficiency and potential search speed improvements\cite{gray1984vector}. Quantixar focuses on two primary quantization techniques:
\begin{itemize}
    \item \textbf{Product Quantization (PQ):}
    
        \textbf{Input:} 
        \begin{itemize}
            \item Original vector: $\mathbf{x} \in \mathbb{R}^d$
            \item Number of sub-vectors: $m$
            \item Codebook size for each sub-vector: $k$
        \end{itemize}
        
        \textbf{Steps:}
        \begin{enumerate}
            \item \textbf{Partitioning:} Divide $\mathbf{x}$ into $m$ sub-vectors: 
            $\mathbf{x} = [\mathbf{x}^{(1)}, \mathbf{x}^{(2)}, ..., \mathbf{x}^{(m)}]$
            
            \item \textbf{Codebook Construction:} For each sub-space, learn a codebook:
            $\mathcal{C}^{(i)} = \{\mathbf{c}_1^{(i)}, \mathbf{c}_2^{(i)}, ..., \mathbf{c}_k^{(i)}\} \subset \mathbb{R}^{d/m}$
            
            \item \textbf{Encoding:} Quantize each sub-vector by finding its nearest centroid:
            \[ q(\mathbf{x}^{(i)}) = \arg\min_{\mathbf{c} \in \mathcal{C}^{(i)}} ||\mathbf{x}^{(i)} - \mathbf{c}||^2 \]
            
            \item \textbf{Quantized Representation:} 
            $\hat{\mathbf{x}} = [q(\mathbf{x}^{(1)}), q(\mathbf{x}^{(2)}), ..., q(\mathbf{x}^{(m)})]$ 
        \end{enumerate}

    \item \textbf{Binary Quantization (BQ):}
    
        \textbf{Input:} 
        \begin{itemize}
            \item Original vector: $\mathbf{x} \in \mathbb{R}^d$
            \item Number of hyperplanes: $m$
        \end{itemize}
        
        \textbf{Steps:}
        \begin{enumerate}
            \item \textbf{Hyperplane Learning:} Learn a set of hyperplanes with normal vectors $\mathbf{u}_1, \mathbf{u}_2, ... \mathbf{u}_m \in \mathbb{R}^d$ 
            
            \item \textbf{Encoding:} Generate a binary code $\mathbf{b} = [b_1, b_2, ..., b_m]$ where:
            \[ 
            b_i = 
            \begin{cases}
                1, & \text{if } \mathbf{u}_i^T \mathbf{x}  \geq 0 \\
                0, & \text{otherwise} 
            \end{cases}
            \]
            
            \item \textbf{Binary Representation:} The binary code $\mathbf{b}$ is the compact representation of $\mathbf{x}$
        \end{enumerate}
\end{itemize}

\subsubsection{Advanced Techniques: Harnessing Hardware with SIMD}

Modern processors offer a powerful tool for accelerating the computations at the heart of vector databases: SIMD (Single Instruction, Multiple Data) instruction sets. Found in extensions like SSE (Streaming SIMD Extensions) and AVX (Advanced Vector Extensions), these instructions let the processor perform the same operation on multiple pieces of data simultaneously, offering significant performance gains\cite{8130042, 10458910}.

\begin{itemize}
    \item \textbf{SSE (Streaming SIMD Extensions)}
    \begin{itemize}
        \item \textbf{XMM Registers:} The heart of SSE lies in its introduction of eight 128-bit XMM registers (XMM0 through XMM7)\cite{russinoff2021sse}. These specialized registers are designed to hold multiple packed data elements. For example, each XMM register can store:
        \begin{itemize}
            \item Four 32-bit floating-point numbers
            \item Eight 16-bit integers
            \item Sixteen 8-bit integers
        \end{itemize}
        \item \textbf{Parallelism Within Registers:}  SSE instructions operate on these packed data elements simultaneously. Let us imagine we have two XMM registers containing four 32-bit floating-point numbers each:
        \begin{itemize}
            \item XMM0: [A1, A2, A3, A4]
            \item XMM1: [B1, B2, B3, B4]
            \item A single SSE addition instruction can perform the following in one operation: [A1+B1, A2+B2, A3+B3, A4+B4]
        \end{itemize}
    \end{itemize}
    
    \item \textbf{AVX2 (Advanced Vector Extensions 2)}
    \begin{itemize}
        \item \textbf{YMM Registers:}  AVX2 extends the XMM registers to a wider 256-bit format, creating sixteen YMM registers (YMM0 through YMM15). These larger registers double the potential data processing throughput\cite{gepner2017using, guide2011intel}.
        
        \item \textbf{Fused Multiply-Add (FMA):} AVX2 introduces support for FMA instructions. These instructions combine a multiplication and addition operation into a single step (e.g., D = (A * B) + C) \cite{gepner2017using, guide2011intel}. This type of calculation is common in many vector and matrix operations, providing a performance boost.
        
        \item \textbf{Register Relationship:} It is important to note that YMM registers encompass the XMM registers. The lower 128 bits of any YMM register correspond to the respective XMM register (e.g., the lower half of YMM0 is the same as XMM0). This allows for some flexibility in code that mixes SSE and AVX2 instructions.
    \end{itemize}
\end{itemize}

\section{Methodology}
\begin{figure}[htbp]
    \centerline{\includegraphics[width=0.5\textwidth]{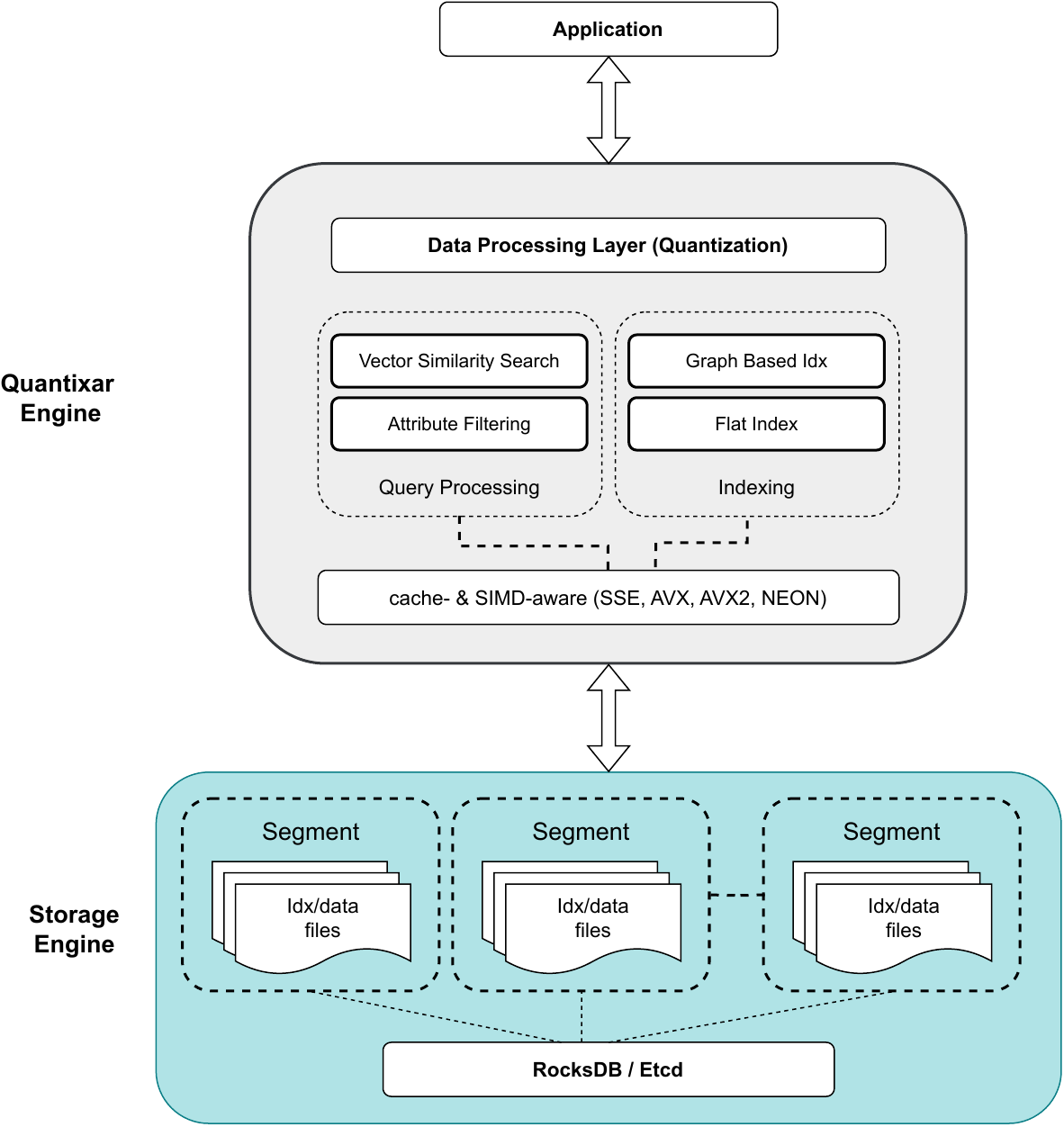}}
    \caption{Quantixar System Architecture.}
    \label{fig}
\end{figure}

This section delves into Quantixar's architectural design, providing a comprehensive overview of its core components and their interplay. Illustrated in Fig.~\ref{fig}, Quantixar's architecture comprises four principal modules: Query processing module, Quantization module, Indexing, and Storage Engine module. Each module is critical to the system's overall functionality and performance.

\subsection{Query Processing :}
In this subsection, we introduce the concept of an ``entity" as it pertains to the Quantixar Engine and discuss the types of queries supported by this engine, the various similarity functions employed, and the application interfaces designed for user interaction.

\textbf{Concept of Entity:}
Within the Quantixar system, an ``entity`` is the basic unit of data representation and is crucial to understanding the database's operation. An entity is typically a high-dimensional vector that serves as an abstract representation of complex data items. These data items, such as images, texts, audio files, or other forms of multimedia, are transformed into vectors through feature extraction methods. The vector encapsulates the key attributes of the original data, enabling efficient execution of operations like similarity searches or clustering within the database.

\textbf{Query Types:} Quantixar supports two primitive query types:

\begin{itemize}
    \item \textit{\textbf{Vector Query:}} 
        The vector query is a foundational query type in Quantixar, primarily focusing on vector similarity search. In this context, each data entity is represented as a single vector. The primary function of a vector query is to identify and retrieve entities that are most similar to a given query vector\cite{rashtchian2020vector}. This process involves computing similarity based on specific metrics or distance measures, determining how closely entities match the query vector. The vector query is critical in scenarios where the similarity between entities, represented as vectors, is the primary criterion for data retrieval.
    \item \textit{\textbf{Metadata-Enhanced Vector Search (MEVS):}}
        It is a query method that augments vector-based searches by incorporating attribute-based constraints. In this approach, each vector is associated with metadata containing key attributes and contextual details\cite{jokela2001metadata}. MEVS applies these attribute constraints to the metadata, effectively narrowing the dataset to only those entities meeting the specified criteria. The system conducts a vector similarity search within this selectively filtered subset. This sequential method, which starts with metadata-based filtering and then proceeds to vector similarity analysis, ensures more efficient and pertinent query results. By focusing the vector search within a refined subset, MEVS achieves faster response times and heightened accuracy in matching\cite{chiu2016minimum,bookstein2002generalized,leydesdorff2008normalization,seal2020fuzzy}.
\end{itemize}

\subsection{Quantization}
The Data Preprocessing / Quantization module in Quantixar is the critical first step, laying the groundwork for efficient similarity search within the high-dimensional vector space. This section focuses on transforming dense vector data into a sparse vector. The configuration provided by the user controls this transformation, which is done using quantization methods, including product and binary quantization.

\subsection{Indexing}
Quantixar's indexing architecture ensures efficient and scalable retrieval of high-dimensional vectors. It offers a dual indexing strategy, HNSW and Flat Index, to accommodate diverse deployment scenarios and performance optimizations.
The choice between HNSW and flat indexing hinges on carefully considering dataset size, desired query speed, and the tolerance for approximation in search results. Flat Indexing, while simple, offers the guarantee of finding the actual exact nearest neighbors to a query vector. This makes it suitable for smaller datasets where computational cost is manageable or in use cases where the utmost precision is non-negotiable. However, as data points grow, the linear search time required for flat Indexing quickly becomes a significant performance bottleneck. Flat Indexing becomes impractical if the application demands scalability to large datasets.
This is where HNSW indexing excels. By constructing a hierarchical graph-based structure on the vector dataset, HNSW enables approximate nearest neighbor (ANN) search with substantially improved efficiency. Search times scale logarithmically with the dataset size, allowing HNSW to handle datasets with orders of magnitude larger than those feasible with flat Indexing. However, it is essential to note the "approximate" nature of the results. HNSW introduces a degree of trade-off between absolute precision and search speed. With careful tuning, HNSW can achieve remarkably high accuracy while dramatically outperforming flat Indexing in speed.

By leveraging these indexing methods, and based on query characteristics and dataset size, the Query Processing module will determine the type of index to utilize and passes the query to the Storage Engine for data retrieval.

\subsection{Vector Storage}
In the storage architecture of Quantixar, both RocksDB and etcd play crucial roles in catering to different user needs. RocksDB, pivotal for vector embedding persistence, was chosen for its high-performance characteristics in key-value storage and is specially optimized for SSDs. Its data structure handles efficient read and write operations, supported by block-size management, thread pool, and advanced compaction strategies\cite{matsunobu2020myrocks, kim2019optimizing}. Moreover, MVCC management at the cluster level makes it ideal for high-speed data processing environments.

Etcd, on the other hand, is integrated for environments requiring distributed data management\cite{larsson2020impact, nalawala2022comprehensive}. etcd's datastore is built on BBoltDB, offering efficient caching via a single memory-mapped file managed by the operating system and an in-memory B-tree index for quick key-revision mapping\cite{nalawala2022comprehensive}. This structure suits distributed systems, efficiently supporting range reads and sequential writing. etcd's key-value store is optimal for scenarios where high availability, consistency, and fault tolerance in a distributed environment are paramount\cite{larsson2020impact}.

Quantixar offers a flexible storage solution, allowing users to opt for RocksDB or etcd based on their specific requirements, whether high-speed data processing in non-distributed environments or robust, distributed data management.

\section{Results}
This experiment evaluated the performance of the Hierarchical Navigable Small World (HNSW) indexing algorithm within the Quantixar framework.  Performance was assessed on two datasets: Fashion-MNIST and SIFT 128. Key metrics are reported in Table ~\ref{table:performance_multi_dataset}, including construction time, insertion time, search time (at different 'ef' values), recall rates, and more.

\begin{table}[htbp]
\centering
\caption{Performance Metrics for HNSW in Quantixar on Various Datasets}
\label{table:performance_multi_dataset}
\begin{tabular}{|l|c|c|}
\hline
\textbf{Metric} & \textbf{Fashion-MNIST} & \textbf{SIFT 128} \\ \hline 
Construction Time (s) & 0.22887 & 26.04492 \\ \hline
Insertion Time (s) & 38.27 & 5655.94 \\ \hline
Search Time (s) (ef=64) & 6.05864 & 130.16474 \\ \hline
Search Time (s) (ef=128) & N/A & 34.84278 \\ \hline
Mean Fraction of Neighbors Returned & 1.0 & 1.0  \\ \hline
Last Distances Ratio (ef=64) & 1.001 & 1.0003 \\ \hline
Last Distances Ratio (ef=128) & N/A & 1.0002 \\ \hline
Recall Rate (ef=64) & 0.978 & 0.9908 \\ \hline
Recall Rate (ef=128) & N/A & 0.9964 \\ \hline
Queries per Second (ef=64) & 10647.42 & 4158.96 \\ \hline
Queries per Second (ef=128) & N/A & 2051.11\\ \hline
\end{tabular}
\end{table}

\textbf{System Specifications:}
AWS EC2 t4g.xlarge instance (4 vCPUs, 16GB RAM)

As expected, construction and insertion time increase for the more complex SIFT 128 dataset.  Notably, the recall rate is excellent across both datasets, even at the lower 'ef' value, indicating HNSW's effectiveness in approximate nearest neighbor search.

\section{Conclusion}
In conclusion, the development of Quantixar reflects our commitment to advancing the field of vector databases by learning from and building upon the foundations laid by established systems like Qdrant, Weaviate, and Milvus. Each of these systems offers unique insights into handling large-scale, high-dimensional data, which have been instrumental in shaping the design and functionality of Quantixar. With its integrated components, such as the Quantixar Engine, Index, and Storage Engine, Quantixar aims to enhance operations like similarity searches and efficient data retrieval. Recognizing this field's dynamic and collaborative nature, our approach is not just about creating a new solution but also about contributing to the collective knowledge and technology base. As we move forward, our focus will be on empirical validation and performance benchmarking, ensuring that Quantixar not only meets but also enriches the evolving standards of vector database technology. Through this ongoing journey, we aim to solidify Quantixar's place as a valuable and innovative addition to high-dimensional data processing.

\vspace{12pt}

\end{document}